# Exploring Large Language Models for Semantic Analysis and Categorization of Android Malware

Brandon J Walton*, Mst Eshita Khatun†, James M Ghawaly‡, Aisha Ali-Gombe§
*Dept. of Computer Science & Engineering, Louisiana State University*
Baton Rouge, United States
Email: *bwalto8@lsu.edu, †mkhatu3@lsu.edu, ‡jghawaly@lsu.edu, §aaligombe@lsu.edu

*Abstract*—Malware analysis is a complex process of examining and evaluating malicious software's functionality, origin, and potential impact. This arduous process typically involves dissecting the software to understand its components, infection vector, propagation mechanism, and payload. Over the years, deep reverse engineering of malware has become increasingly tedious, mainly due to modern malicious codebases' fast evolution and sophistication. Essentially, analysts are tasked with identifying the elusive needle in the haystack within the complexities of zero-day malware, all while under tight time constraints. Thus, in this paper, we explore leveraging Large Language Models (LLMs) for semantic malware analysis to expedite the analysis of known and novel samples. Built on GPT-4o-mini model, *MalParse* is designed to augment malware analysis for Android through a hierarchical-tiered summarization chain and strategic prompt engineering. Additionally, *MalParse* performs malware categorization, distinguishing potential malware from benign applications, thereby saving time during the malware reverse engineering process. Despite not being fine-tuned for Android malware analysis, we demonstrate that through optimized and advanced prompt engineering *MalParse* can achieve up to 77% classification accuracy while providing highly robust summaries at functional, class, and package levels. In addition, leveraging the backward tracing of the summaries from package to function levels allowed us to pinpoint the precise code snippets responsible for malicious behavior.

## 1. Introduction

Reverse engineers face significant challenges in analyzing and understanding malware samples in the ever-evolving cybersecurity domain. Even more so when dealing with suspected novel or zero-day malware, reverse engineering the unknown binary or program file is crucial. This approach involves dissecting the malware's code, structures, and behaviors to understand its mechanisms. However, reverse engineering demands a high skill level and deep knowledge of coding, system architecture, and cybersecurity principles. As such, understanding malware capabilities through reverse engineering can be time-consuming and complex especially for advanced, and obfuscated trojans with extensive code bases such as Android malware. Thus, in this paper, we propose a technique to enhance the current static process of Android malware analysis by exploring the capabilities of pre-trained Large Language Models (LLMs). Our proposed approach, titled *MalParse* , allows for categorizing malware, detailed code summarization, and efficient backtracking within the codebase to identify the root causes of malicious activity. Essentially, *MalParse* streamlines malware analysis by providing analysts with a summary of the malware and a direct path to its root causes. To achieve this, *MalParse* is designed with three major components: Auto-decompilation and Feature Extraction, Prompt Engineering, and Hierarchical-Tiered Code Summarization modules. In the Auto-decompilation and Feature Extraction module, a target Android application is reverse-engineered and decompiled to access its source code. The proposed Hierarchical-Tiered Summarization module breaks down and summarizes extracted code structures at multiple granularity levels, specifically focusing on function-level, class-level, and package-level summaries. Starting from the most basic building blocks of a program, individual functions are summarized, and these summaries are then aggregated and re-summarized to generate contextual summaries for each class. This iterative process is applied to all classes within the Android application. Finally, the class summaries are consolidated and fed into the model to generate a comprehensive package summary, providing an overview of the application's functionality, identifying any detected malicious activities, if present, and tagging the sample as either malicious or benign. Furthermore, our proposed approach is further tuned using Prompt Engineering to improve the accuracy of a foundational LLM, which was not fine-tuned on malware. This component of *MalParse* pre-instructs the model with three prompt categories to guide the model's understanding of malware: Vanilla for functionality overview, API-Scoped for common permissions and APIs used in malware, and Malware-Scoped for identifying malicious indicators such as privilege escalation and dynamic code execution.

Evaluating the effectiveness of our approach, *MalParse* was tested on a dataset comprising 200 Android applications, split evenly between malware and benign. The opti-mized *MalParse* model, which is based on Malware-Scoped prompting, achieved a 77% categorization rate in accurately identifying whether an Android application is benign or malicious, outperforming the Vanilla prompts' 49.5% and the API-Scoped prompts' 56%. This highlights the inadequacy of the naive prompting approach using the Vanilla prompts for malware summarization. Importantly, it is worth noting that the model was intentionally not pre-trained on previous malware or benign samples. This emphasizes the significance of the 77% categorization rate; our general LLM (which was not built with code summarization objective function) achieved this accuracy without prior training on any sample, simulating the approach of analyzing zero-day malware. Additionally, the only requirement is the expert's knowledge, which is simulated by prompt engineering. In addition, we also conducted efficient backtracking through the summaries, spanning from package to class and function levels, to identify the root cause of the summary decision chain of thought and malware functionality. The result of this evaluations further proved *MalParse* 's efficacy and correctness in itemizing program functionality.

**Contributions -** This work makes the following salient contributions:

- Comprehensive Malware Summaries: The hierarchical-tiered summarization approach of *MalParse* at different levels (function, class, and package) ensures that the final outputs are not only comprehensive but also contextually relevant to the security analysis, providing precise and actionable insights.
- Accuracy: By fine-tuning *MalParse* using Prompt Engineering to better understand and recognize patterns associated with malware, the research serves as a baseline that shows how naive LLMs can significantly improve their accuracy in malware analysis tasks, even if they are not pre-trained on any samples.
- Efficacy: Through the process of backtracking and chain of thought analysis, *MalParse* provides a sound and reliable semantic analysis system for reverse engineers to not only understand the high-level application description but also explore the root cause for program functionalities in the code base as well as explore the summary decisions of *MalParse* .

**Paper outline.** The rest of the paper is structured as follows, Section 2 provide the review of related literature in code summarization. Section 3 outlines the design and implementation of the proposed *MalParse* . Section 4 presents the evaluation of *MalParse* , its limitations and future work. Finally, Section 5 summarizes our findings and conclusions.

## 2. Literature Review

Automated code summerization are techniques that generates natural language descriptions of high-level source code using machine learning. From a security standpoint, integrating code summarization techniques with malware analysis signifies a substantial advancement in cybersecurity initiatives aimed at analyzing malware threats.

### 2.1. LLMs for Code Summarization

Recent advancements in generative AI, notable Large Language Models (LLMs), have sparked a growing interest in the use of LLMs for tasks involving the understanding and generation of code documentation, repair, and testing. One example of an LLM developed for such tasks is CodeBERT [1], which is trained on both natural and programming languages aimed at improving code search and comprehension. Similarly, GraphCodeBERT [2] integrates structural code information into the CodeBERT training process to enhance its performance on code related tasks. PLBART [3] on the other hand explores the utility of LLMs trained in a denoising autoencoding paradigm for a variety of programming tasks, including code documentation and summarization. Large foundational LLMs have also garnered significant attention for programming tasks. OpenAI's GPT-3 [4], has demonstrated exceptional abilities in producing text, including coherent code segments. Many studies [5], [6], [7], [8], [9] have evaluated ChatGPT's automatic bug resolution performance, revealing its potential nature for enhancing software repair methodologies. Zahan et al. [10] investigated a variety of LLMs, including GPT-3 and GPT-4, for the purpose of detecting malware in the npm ecosystem (JavaScript code). Their study served as a baseline for JavaScript malware analysis and demonstrated notable improvements over static analysis tools in terms of precision and F1 scores. In contrast, our proposed *MalParse* explores Android malware analysis using a hierarchical-tiered approach with function-, class-, and package-level summaries. The novelty of our work lies not only in the categorization of malware, which we believe will be highly beneficial for analysts, but also in the final package-level descriptions aimed at aiding analysts in evaluating zero-day samples. Our evaluation demonstrates the critical impact of strategic and effective prompt engineering in enabling the LLM to effectively categorize and generate enhanced descriptions of target samples.

## 3. Design and Implementation

This section outlines the conceptual methodology and implementation of the proposed Malware Semantic Parser - *MalParse* . Broadly, this approach can be described around three primary steps: 1) auto-decompilation and static feature extraction, 2) hierarchical-tiered summarization, and 3) model optimization using advanced prompt engineering techniques to contextualize the summarization module for APK package analysis.

### 3.1. Auto-decompilation and Feature Extraction Module

The initial step in analyzing any program of unknown origin involves reverse engineering, typically accomplished through disassembly, decompilation, or a combination of both [11]. Since Android applications are developed in Java and compiled into a compressed intermediate format known as Davik Bytecode (Dex), decompilation is the standard technique for these apps. In our decompilation workflow, we create an automated Python-based tools that decomples a target APK file using Dex2Jar [12] to generate Java bytecode. This is then further decompiled using CFR [13] to translate the recovered bytecode into human-readable source code.

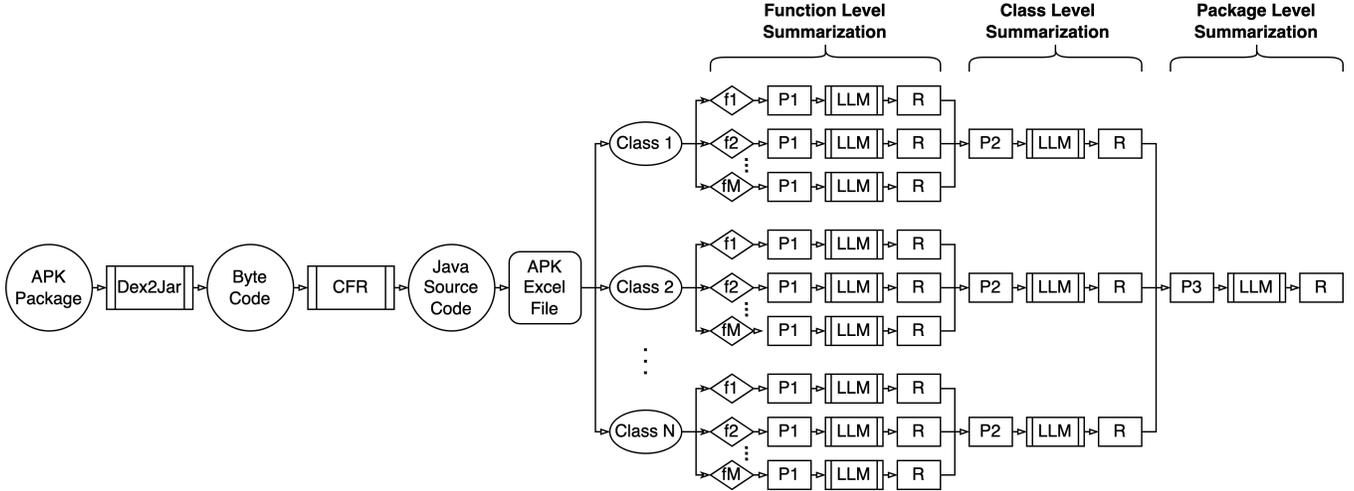

Figure 1. Diagram of the *MalParse* workflow outlining the hierarchical-tiered Code summarization process. It starts with function-level summarization for each method of each class in the package, followed by class-level summarization, and finishing with package-level summarization. *P* refers to a prompt, *R* refers to an LLM response, and *f* refers to a class method.

### 3.2. Hierarchical-Tiered Code Summarization

Following the code extraction, our approach leverages a hierarchical-tiered code summarization to conduct a thorough analysis of the program files and functions as shown in Figure 1. The summarization module which was implemented in Python and compatible with any general-purpose LLM priotize the identification and concise summarization of potential program functionality inherent in the code base. It does so using a bottom-up iterative approach, beginning with summarization at the individual function level and working up to the full package-level. The hierarchical approach offers two key advantages. First, it allows the use of LLMs with restricted context windows, resulting in cost savings and reduced resource requirements. Second, it generally produces superior summarization quality for large contexts compared to a single-step approach [14], [15]. In designing *MalParse* , we focus on OpenAI's GPT-4o-mini model due to its high-performance on major benchmarks, low cost, and its large 128k token context window, which enables the processing of larger chunks of code. We also use the Langchain [16] library for implementing custom LLM chains and prompt templates.

### 3.3. Model Optimization using Advanced Prompt Engineering

In *MalParse* , during the code analysis and summarization, three sets of prompting are utilized - Vanilla, API-scoped, and Malware-scoped. Each category provides the LLM with different contextual information and scope on its objective.

- In the Vanila-scoped, the LLM is not provided with any information about what constitutes potentially suspicious features.
- In the API-scoped, the prompts provide the LLM with information on identifying key sensitive APIs, permissions, and libraries commonly found in Android malware.
- In the Malware-Scoped, the prompts offer the model additional information for identifying malicious functionalities and malware types such as dynamic class loading, rooting, data exfiltration etc.

Each of these prompts enable the model to dynamically expand its context window by iteratively summarizing all functions, thereby generating class-level summaries. Subsequently, these class-level summaries are combined to produce package-level summaries. This hierarchical approach results in a more comprehensive contextual summary of potential security threats and common activities within the codebase. To craft the most effective prompts, several techniques were employed:

**1. Few-Shot Prompting:** The technique of few-shot prompting was employed, where expected input and output formats were provided to guide the model. This approach ensured that the model received clear instructions on the desired input structure and the expected format for generated summaries, facilitating more accurate and relevant output.

**2. Context Injection:** Through tailored prompts (API- and Malware-Scoped), contextual cues are injected to guide the LLM in summarizing code snippets, with a particular emphasis on dangerous APIs, sensitive permissions, and unusual programming practices present in the code. This approach ensures that the final summary retains pertinent context, specifically addressing security considerations.

**3. Integration of External Knowledge:** In the package-level summarization for API- and Malware-scoped, the prompts were strategically modified to incorporate insights from external sources, including knowledge of known malware and common activities observed in malware applications. This proactive approach effectively guides the model in accurately identifying potential security threats and malicious behavior.

This conceptual design and implementation lays a general foundation for semantic malware analysis leveraging LLMs, which could be adapted for other classes of software beyond Android malware analysis.

## 4. Evaluation

In this section, we evaluate the performance of *MalParse* based on two criteria: 1) **Accuracy** in distinguishing malware from benign Android applications across three different prompting categories; 2) the **Efficacy** of the model's response after package-level summarization.

### 4.1. Experimental Setup

**Dataset and Preprocessing:** To ensure reproducibility, our dataset comprised 200 APK files, with 100 identified as known malicious applications and the remaining 100 as known benign applications. The malicious samples, sourced from VirusTotal, included various malware categories such as ransomware, scareware, adware, and spyware. Notable examples within these categories included WannaLocker, MazarBot, and Jifake. The benign samples, sourced from the top 500 applications listed by Similarweb in the United States[17], included notable apps such as Facebook, Google, YouTube, LinkedIn, Outlook, and Instagram. These benign samples served as a reliable benchmark to assess the LLM's ability to differentiate between malicious and benign APK files. Each APK sample is processed through the auto-decompilation and feature extraction process as described in Section 3, resulting in the decompilation of all classes within each APK and their preparation for input. Additionally, to mitigate bias in the LLM's summaries, we did not fine-tune (train) the LLM on any known examples of benign or malicious Android applications. This allows the *MalParse* to conduct semantic analysis on "newly" encountered APKs without any contextual bias. **This approach is critically important as it simulates an analyst examining a 0-day sample for the first time without prior knowledge of what to expect.** Furthermore, to mitigate the GPT-4o-mini model from relying on memory, the names of the APKs were not provided. This aids in reducing the model's reliance on external information such as the name of the application, aiding in the prevention of it categorizing the application before deeper code analysis.

**Model Execution:** After pre-processing, the input features for each APK are passed to the Code Summarization Module of *MalParse* for execution as detailed in Section 3. To facilitate a more efficient analysis, separate instances of the program were deployed for analyzing the malware and benign samples. Both instances ran concurrently on a local server, allowing simultaneous and independent evaluations of each sample type.

### 4.2. Experimental Results

**4.2.1. R1 - Evaluation of the Categorization Accuracy.** The objective of this first test is to assess how effectively *MalParse* can identify and categorize applications based on their security status (malicious vs. benign) under varied contextual prompts. This metric is crucial, particularly in determining how prompt scoping impacts and influences the general performance of the model.

**1. Vanilla-Based App Categorization:** The Vanilla Prompts are used to provide summaries of the APK's general functionality without referencing any context to what suspicious activities, device permissions, or API calls to observe. The result of our analysis indicates that the Vanilla prompts performed the worst in categorizing the APK's true nature. Examining the confusion matrix in Table 1, we see that 92% of the benign samples (TP) and 7% of the malicious samples (TN) were correctly categorized. Dividing the sum by the total samples (200) yields a **categorization accuracy of only 49.5%**. This indicates these prompts only correctly categorized 99 out of the 200 samples. Overall, the Vanilla prompts demonstrated poor classification accuracy, clearly evidenced by high performance in benign categorization with low performance in malware categorization. Moreover, the only identified malware samples were applications associated with tracking user activity.

| Vanilla Confusion Matrix | | |
|---|---|---|
| Model | Benign (P) | Malware (N) |
| Benign Actual (P) | **92% (TP)** | 93% (FN) |
| Malware Actual (N) | 8% (FP) | **7% (TN)** |

TABLE 1: Confusion Matrix for the Vanilla Prompts

**2. API-Scoping-Based App Categorization.** These prompts provide a list of the common APIs, libraries, and permissions used in Android malware. Unlike the Vanilla prompts, the model is now provided with context regarding the drivers of malicious functionalities to be aware of, aiding in the categorization assessment. Our analysis result showed that the API-scoped prompts ranked as the second-best in correctly categorizing the APK samples. Upon examining the confusion matrix as shown in Table 2, it's apparent that these prompts accurately identified 90% of the benign samples (TP) and 22% of the malware samples. Summing the percentages of correctly identified samples and dividing by the total number of samples (200) resulted in a **categorization accuracy of 56%**. This indicates that the API prompts correctly identified 112 samples out of the 200, marking a higher result than Vanilla's 45% categorization accuracy. The API-Scoped prompts demonstrated superior classification performance compared to the Vanilla prompts. However, this improvement comes with significant trade-offs. Our manual analysis of the summaries revealed that the model displayed a bias towards obfuscation APIs like Java reflection APIs which alters the way in which functions are called. It tends to classify all APKs utilizing reflection as malicious.

| API-Scoped Confusion Matrix | | |
| --- | --- | --- |
| Model | Benign (P) | Malware (N) |
| Benign Actual (P) | **90% (TP)** | 78% (FN) |
| Malware Actual (N) | 10%(FP) | **22% (TN)** |

TABLE 2: Confusion Matrix for the API-Scoped Prompts

**3. Malware-Based Scoping Categorization:** These last prompts provide a detailed outline of the suspicious activities and unusual behaviors commonly found in malware, equipping the model with essential context that will enhance its classification and summarization abilities. Our analysis revealed that Malware-Scoped prompts performed the best out of the three in correctly categorizing the nature of APK samples. As shown in the confusion matrix in Table 3, this prompt scoping correctly classified 76% of benign samples and 78% of malware samples. By summing these percentages and dividing them by the total number of samples (200), the Malware-Scoped prompts achieved a classification rate of 77%, resulting in 154 out of the 200 samples being correctly classified. This percentage significantly surpasses the Vanilla prompts rate of 45.5% and API-Scoped prompts of 56%. Although the Malware-Scoped prompt demonstrated superior classification and lower misclassification rates compared to both the Vanilla and API-Scoped prompts, its misclassification rate of 22% is not ideal. Notably, the misclassification results did not show any bias towards specific categories of suspicious activities mentioned in the prompt. This indicates that certain other suspicious activities, not explicitly covered in the prompts, may have been present in the code base but were not picked by the model due to the lack of guidance in identifying those categories. We strongly believe that expanding the malware prompt to encompass a broader range of malware functionalities will significantly enhance classification accuracy.

| Malware Scoped Confusion Matrix | | |
| --- | --- | --- |
| Model | Benign (P) | Malware (N) |
| Benign Actual (P) | **76% (TP)** | 22% (FN) |
| Malware Actual (N) | 24% (FP) | **78% (TN)** |

TABLE 3: Confusion Matrix for the Malware-Scoped Prompt

**4.2.2. R2 - Evaluation of the Summary Efficacy.** In this evaluation objective, we explore the efficacy and usefulness of the summarization for an analyst. The primary goal is to assess if the summary is: 1) Comprehensive enough to provide adequate information to the analyst; 2) Capable of tracing the chain-of-thought through the prompt decision, enabling the analyst to understand how decisions about malware functionality are reached; and 3) Able to link all elements together to precisely pinpoint the part of the code responsible for that decision. To illustrate this, we present an example in Table 5, showing that the sample

```java
public static void initRoot(Context context) {
    String string = RTUtils.a(context);
    new Thread(new b(context, string, string)).start(); //Calls b
}

static /* synthetic */ void b(Context context, String string, String string2)
{
try {
    Object object;
    if (a == null) {DexClassLoader dexClassLoader;
        object = new StringBuilder(String.valueOf(context.getFilesDir().
        getAbsolutePath()));
        String string3 = ((StringBuilder)object).append(File.separator).append(
        RootFileName).toString();
        object = context.getFilesDir().getAbsolutePath();
        StringBuilder stringBuilder = new StringBuilder(String.valueOf(context.
        getApplicationInfo().
        nativeLibraryDir));
        a = dexClassLoader = new DexClassLoader(string3, (String)object,
        stringBuilder.append("/").toString(),
        ClassLoader.getSystemClassLoader());}
    object = a.loadClass("cn.engine.RootPermApi");
    Object t = ((Class)object).getConstructor(new Class[0]).newInstance(new Object
        [0]);
    object = ((Class)object).getMethod("initRoot", Context.class, String.class,
        String.class);
    ((AccessibleObject)object).setAccessible(true);
    ((Method)object).invoke(t, context, string, string2);
    return;
}
catch (Exception exception) {
    exception.printStackTrace();
    return;
}
}
```

TABLE 4: The code snippet for the initRoot() and b() methods of the RTUtils (RTAccessHandler) class

with hash 9324376e27f9e1ddd05d181d656c6b76 was categorized as malware based on evidence of device rooting found in the class *RTAccessHandler*. The specific package-level prompt questioned whether the app could engage in suspicious activities such as dynamic code execution and privilege escalation. Linking this package-level summary to the *RTAccessHandler* class-level summary (originally called *RTUtils* but renamed by *MalParse* to enhance readability), it was revealed that this class contains two critical functions: *initRoot()* and *executeRoot()*. Further analysis of the class-level summary with the function-level summary showed that *initRoot()* calls the *b()* function, detailed in the Code Listing in Table 4, to load and execute the dynamic class *cm.engine.RootPermAPi*. Another instance of *initRoot* in this class is executed using **Java Reflection**. This evaluation not only validates the summary's comprehensiveness through chain-of-thought and explainability but also shows its capability to provide detailed insights into the nature of the malware, demonstrating the efficacy of *MalParse*.

### 4.3. Discussion

The effectiveness of *MalParse* demonstrated in the evaluation indicates its ability to accurately and effectively summarize and categorize malware. One of the significant capabilities of *MalParse* is its use of Chain-Of-Thought processes, which not only enhance the depth of analysis but also improve the traceability of decisions made by the model. This feature is particularly useful in the context of malware analysis and reverse engineering, where understanding the how and why of the malware, including the step-by-step reasoning of its design, is as important as the payload itself. However, *MalParse* is not without limitations, as shown below.

| *MalParse* Summaries | Prompt Chain-of-Thought and Explainability |
|---|---|
| **Package-level Summary:** DownloadManagerService: This Java package focuses on managing download operations and notifications within the application, handling various background tasks and notification updates. The DownloadingServiceHandler class orchestrates the processing of incoming messages, logging message details, and managing the download queue. While seemingly benign, the **RTAccessHandler class** stands out for managing root access operations and system modifications, executing root commands, and dynamically loading classes for root-related actions. These actions align with Privilege Escalation and Control and Stealth and Resource Exploitation, raising significant security concerns regarding unauthorized system modifications and elevated privilege abuse. The presence of these activities categorizes this package as ⟨MALWARE⟩⟨**Rooting**⟩; ⟨Stealth⟩ and ⟨Resource Exploitation⟩ | **Tag ⟨Rooting⟩:** Gaining privileged access to the root user account on a device or system, typically to bypass limitations imposed by the manufacturer or carrier. A) Related Suspicious Activities: 1. Root Access 2. Modification of Critical System Components 3. Dynamic Code Execution B) Related Category: Privilege Escalation and Control C) Known Examples: Towelroot, KingRoot, Magisk |
| **Class-level Summary:** (RTUtils -)RTAccessHandler): Manages root access operations and system modifications within an Android application. This class includes methods for executing root commands, initializing root access, and dynamically loading classes for root-related actions. The **executeRoot()** function initiates a new thread to execute root commands, while the **initRoot()** method initializes root access by obtaining a string value. These operations involve reflection to interact with hidden APIs and execute root-related tasks, requiring elevated permissions on the device. The class poses potential security risks due to its ability to modify system settings and execute privileged actions. ⟨**Privilege Escalation and Control**⟩;⟨Steal and Resource Exploitation⟩; | **Tag: ⟨Privilege Escalation and Control⟩** 1. Root Access: Attainment of privileged control over a system, often granting unrestricted access to its resources and functions. |
| **Function-level Summary:** initRoot(Context context): Initiates a root-related operation by generating a string using RTUtils and starting a new thread with a specific context. The function appears to be involved in initializing root-related functionalities without explicitly showing any suspicious activities. **executeRoot(Context context, String string, boolean bl):** Executes a function that creates a new thread, passing the context, a string, and a boolean value. The thread starts running a method from class 'c' with the provided parameters.; **b(Context context, String string, String string2):** This function dynamically loads a class "cn.engine.RootPermApi" using a DexClassLoader to execute the "initRoot" method. It initializes root access with parameters passed as context, string, and string2. The suspicious activity present includes dynamic code execution through loading external classes at runtime, potentially for malicious purposes. ⟨**Dynamic Code Execution**⟩ | **Tag: ⟨Code Execution Manipulation⟩** SUSPICIOUS ACTIVITIES: 1. Dynamic Code Execution: Malicious execution of code at runtime to evade detection and perform harmful activities. 2. Obfuscated Code: The deliberate concealment of program logic through intricate transformations to impede scrutiny and identification. 3. Modification of Critical System Components: Unauthorized alterations to essential elements of the system infrastructure, jeopardizing system integrity, stability, and security. |

TABLE 5: Description for the Malware 9324376e27f9e1ddd05d181d656c6b76

**4.3.1. Limitations.** The quality and accuracy of the *MalParse* 's summarization and categorization heavily depend on the underlying GPT-4o-mini and the robustness of the prompt engineering, as was seen with the results in R1. The model's accuracy in categorizing malware and becoming less biased toward benign applications improves as the prompting is tuned to understand the general behavior of Android malware. Additionally, for stealthy and obfuscated malware samples, *MalParse* 's reliance on static summarization will result in less comprehensive information especially for obfuscated malware. Code obfuscation techniques will make it difficult for static summarization approaches to capture and describe all aspects of a malware's functionality. Furthermore, the accuracy of the static decompilation relied upon the accuracy of Dex2Jar and CFR which in some cases doesn't decompile complex classes. It was evident that this resulted in some of the malware samples in our dataset being categorized as benign. Finally, although *MalParse* achieved a balanced accuracy at 77% with a precision of 76% and a recall rate of 77% all without prior training. This being said, the 24% and 22% False Positive and False Negative rates, respectively, can have serious implications in real analysis, either by overlooking actual threats or by causing unnecessary alarm and resource expenditure on benign entities.

**4.3.2. Future Work.** To address these limitations and as part of our future work, first, we aim to improve the prompt engineering to refine how the model processes and responds to input, aiming to improve accuracy and bias. Second, we plan to build a locally-hosted model based on a smaller open-source LLM, which will increase processing speed and address rate limitations. Additionally, we intend to redesign *MalParse* to summarize bytecode directly instead of relying on decompiled Java code extracted via memory analysis. This shift will enable a more direct and granular analysis of APK files, potentially increasing *MalParse* accuracy. This new approach will be tested on large Android APK sample sets.

# 5. Conclusion

This paper introduces *MalParse* a system that leverages the power of LLMs for semantic analysis of Android applications. *MalParse* is built on a hierarchical-tiered code summarization approach and advanced prompt engineering to enhance malware classification and the understanding of its structure and behavior. Our approach produces precise, contextually relevant summaries that emphasize security-critical aspects of the code. In its evaluations, *MalParse* classified malware with 77% accuracy without prior training and generated summaries similar to those from automated engines. The backtracking process effectively traces package-level summaries to function-level summaries, providing a detailed description of the malicious behavior's root cause and an explanation of the model's decision.